\documentclass[12pt]{article}
\usepackage{amsfonts}
\usepackage{verbatim}
\usepackage{graphicx}
\usepackage{amsmath}
\usepackage{amssymb}

\begin{document}

\begin{center}
 
{\bf {\large Jackiw-Pi Model: A Superfield Approach\footnote{Talk delivered at BLTP, JINR, Dubna (Moscow Region), Russia 
in the International Workshop on  ``Supersymmetries \& Quantum Symmetries''(SQS'2013) during July 29 - August 3, 2013.} }}

\vskip 1 cm

{\bf Saurabh Gupta} \\
{{\it The Institute of Mathematical Sciences, CIT Campus, Chennai - 600 113, India}}\\
{\small {\bf e-mails: saurabh@imsc.res.in, guptasaurabh4u@gmail.com}}
\end{center}
\noindent
{\bf Abstract:} We derive the off-shell nilpotent and absolutely anticommuting Becchi-Rouet-Stora-Tyutin (BRST) as well as anti-BRST 
transformations $s_{(a)b}$ corresponding to the Yang-Mills gauge transformations of 3D Jackiw-Pi model by exploiting the ``augmented" 
superfield formalism. We also show that the Curci-Ferrari restriction, which is a hallmark of any non-Abelian 1-form gauge theories, 
emerges naturally within this formalism and plays an instrumental role in providing the proof of absolute anticommutativity of $s_{(a)b}.$ \\

\vskip 0.25 cm



\section{Introduction} 
Standard Model (SM) of particle physics accounts for three out of four fundamental interactions of nature. 
In spite of the stunning success of SM, which is based on the non-Abelian 1-form gauge theories, one of the main issues 
with gauge theories are connected with the co-existence of mass and gauge invariance together. However, the gauge invariance 
does not necessarily imply the masslessness of gauge particles for sufficiently strong vector couplings \cite{sch}. 
In this context, it is worth mentioning about the models where 1-form gauge field acquires a mass in a natural fashion such as 
4D topologically massive (non-)Abelian gauge theories (with $B \wedge F$ term) \cite{fre,aln,srm}. But, these models suffer from the problems 
related with renormalizability, consistency and unitarity. 

Furthermore, massive gauge theories in other than $(3 + 1)$-dimensions of spacetime, which are free from the problems of 4D topologically massive models, 
have been studied for a quite some time (see, e.g. \cite{des}). The $(2 + 1)$-dimensional Jackiw-Pi (JP) model is one such model where mass and 
gauge-invariance exist together. The JP model is a parity even model and endowed with two sets of local continuous symmetries, namely; the usual Yang-Mills (YM) and
non-Yang-Mills (NYM) symmetries. This model has been studied throughly (see, e.g. \cite{day,del,srm1,sgrk}). 

In this write-up, we have applied ``augmented'' superfield approach to BRST formalism in order to derive the off-shell nilpotent and absolutely anticommunting 
(anti-)BRST symmetry transformations corresponding to the YM gauge symmetry transformations of JP model. The anticommutativity of 
(anti-)BRST symmetry transformations is ensured by the Curci-Ferrari (CF) restriction which emerges naturally in this framework. 
We would like to point out that, within the framework of superfield formalism, a general set up for BRST 
analysis of a general gauge system also exists \cite{lov}. Our present analysis could be thought of as an application of this approach to a 
specific model having a closed gauge algebra. 

\section{Jackiw-Pi Model: Symmetries} 
The Lagrangian density of $(2 + 1)$-dimensional Jackiw-Pi model\footnote{Here we take the 3D flat Minkowski metric 
$\eta_{\mu\nu} = diag (-1, +1, +1)$ and the Levi-Civita tensor follows $\varepsilon_{\mu\nu\eta} \varepsilon^{\mu\nu\eta} = - 3!, \; 
\varepsilon_{\mu\nu\eta} \varepsilon^{\mu\nu\sigma} = - 2! \; \delta^\sigma_\eta,$ etc., with $\varepsilon_{012} = +1 = 
- \varepsilon^{012}$. We adopt dot and cross products $R \cdot S = R^a S^a, \; R \times S = f^{abc} R^a S^b T^c$ 
in the $SU(N)$ Lie algebraic space spanned by the generators $T^a$ satisfying the algebra $[T^a, T^b] = f^{abc} T^c$ with $a,b,c... = 1,2,3,..., N^2-1$.
The covariant derivative is defined as $D_\mu B^a = \partial_\mu B^a - g (A_\mu \times B)^a$. } 
is given as follows \cite{del, srm1}:
\begin{eqnarray}
{\cal L} = - \frac{1}{4} F^{\mu\nu} \cdot F_{\mu\nu} - \frac{1}{4} \big(G^{\mu\nu} 
+ g\; F^{\mu\nu} \times \rho\big) \cdot \big(G_{\mu\nu} + g \;F_{\mu\nu} \times \rho\big) 
+ \frac {m}{2} \; \varepsilon^{\mu\nu\eta} \; F_{\mu\nu} \cdot \phi_\eta, \label{l1}
\end{eqnarray}
where $F_{\mu\nu} = \partial_\mu A_\nu - \partial_\nu A_\mu - g(A_\mu \times A_\nu)$ and  $G_{\mu\nu} = D_\mu \phi_\nu - D_\nu \phi_\mu$ 
are 2-form curvature tensors corresponding to the 1-form fields $A_\mu$ and $\phi_\mu$, respectively. 
Moreover, $\rho$  is a scalar field and $m$ represents the mass parameter.  
In the above, $A_\mu$ and $\phi_\mu$ have opposite parity which makes JP model to be a parity conserving model.

The above Lagrangian density (\ref{l1}) respects two sets of local symmetry transformations, the YM gauge transformations ($\delta_1$) and NYM 
gauge transformations ($\delta_2$), namely \cite{del, srm1};
\begin{eqnarray}
&&\delta_1 A_\mu = D_\mu \Lambda, \qquad \delta_1 \phi_\mu = - g\;(\phi_\mu \times \Lambda), 
\qquad \delta_1 \rho = - g\;(\rho \times \Lambda), \label{l2}
\end{eqnarray}
\begin{eqnarray}
\delta_2 A_\mu = 0, \qquad \delta_2 \phi_\mu = D_\mu \Sigma  , \qquad \delta_2 \rho 
= +\; \Sigma, \qquad \delta_2 F_{\mu\nu} = 0, \label{l3}
\end{eqnarray}
where $\Lambda = \Lambda \cdot T$ and $\Sigma = \Sigma \cdot T$ are $SU(N)$ valued infinitesimal gauge parameters corresponding 
to YM and NYM gauge transformations, respectively. It is straightforward to check that $\delta_1$ and $\delta_2$ are the symmetry 
transformations, as: $\delta_1 {\cal L} = 0, \quad \delta_2 {\cal L} = \partial_\mu \Big[\frac {m}{2}\; 
\varepsilon^{\mu\nu\eta} \;F_{\nu\eta} \cdot \Sigma \Big].$

\section{Augmented Superfield Approach: A Synopsis}
We apply Bonora-Tonin's superfield formalism \cite{bon} to derive the off-shell nilpotent and absolutely anticommuting (anti-)BRST symmetry 
transformations corresponding to the YM symmetries of the JP model. For this purpose, we first generalize the 3D basic fields to their corresponding 
superfields on the $(3,2)$-dimensional supermanifold parametrized by superspace variables $Z^M = (x^\mu, \theta, \bar \theta)$ where $x^\mu$  ($\mu = 0,1,2$) 
are spacetime variables and $\theta, \bar\theta$ are Grassmannian variables (with $\theta^2 = \bar \theta^2 = 0, \; \theta \,\bar \theta + \bar \theta\, \theta = 0$). 
We also generalize the ordinary 3D exterior derivative ($d$) to $(3, 2)$-dimensional superexterior derivative ($\tilde d$). The explicit expressions are as follows:
\begin{eqnarray}
&& A_\mu (x) \longrightarrow \tilde B_\mu (x, \theta, \bar\theta), \qquad C (x) \longrightarrow \tilde F (x, \theta, \bar\theta), \qquad \bar C(x) \longrightarrow \tilde{\bar F} (x, \theta, \bar\theta), \nonumber\\
&&  A^{(1)} \longrightarrow   \tilde A^{(1)} = dZ^M \tilde A_M \equiv dx^\mu \;\tilde B_\mu (x,\theta,\bar\theta) + d \theta\; \tilde {\bar F} (x,\theta,\bar\theta) + d \bar \theta \;\tilde F (x,\theta,\bar\theta), \nonumber\\
&& d \longrightarrow  \tilde d = dZ^M \; \partial_M  \equiv dx^\mu \;\partial_\mu + d \theta \;\partial_\theta + d \bar \theta \;\partial_{\bar\theta}. \label{l4}
\end{eqnarray}
Here, $\tilde B_\mu (x, \theta, \bar \theta), \tilde F(x, \theta, \bar \theta)$ and 
$\tilde {\bar F}(x, \theta, \bar \theta)$ are the superfields on the (3, 2)-dimensional 
supermanifold and $\partial_M = (\partial_\mu, \partial_\theta, \partial_{\bar \theta})$. 
In the second step, these superfields are expanded along Grassmannian direction ($\theta, \bar\theta$) as
\begin{eqnarray}
\tilde B_{\mu} (x, \theta, \bar\theta) &=& A_\mu (x) + \theta\; \bar R_\mu (x) + \bar \theta\; R_\mu (x)
+ i \;\theta \;\bar\theta \; S_\mu (x), \nonumber\\
\tilde F (x, \theta, \bar\theta) &=& C (x) + i\;\theta\; \bar B_1 (x) + i\;\bar \theta\; B_1 (x)
+ i \;\theta\; \bar\theta \; s (x), \nonumber\\
\tilde {\bar F} (x, \theta, \bar\theta) &=& \bar C (x) +  i\;\theta\; \bar B_2 (x) + i\;\bar \theta\; 
B_2 (x) + i \;\theta \;\bar\theta \; \bar s (x), \label{l5}
\end{eqnarray}
where, $R_\mu (x), \bar R_\mu (x), s(x), \bar s(x)$ are fermionic secondary fields and $S_\mu (x), B_1 (x), \bar B_1(x), B_2 (x),$ $ \bar B_2 (x) $ 
are bosonic in nature. Finally, we take the help of horizontality condition (HC) to determine the relationship amongst the basic and secondary fields 
of the theory. 

We note that the kinetic term corresponding to the gauge field $A_\mu$ remains invariant under the gauge transformations (\ref{l2}). Thus, the HC 
implies that it should not be affected by the presence of Grassmannian variables when generalized onto the $(3, 2)$-dimensional supermanifold. The above 
statement can be, mathematically, expressed as  
\begin{eqnarray}
- \frac {1}{4} \; F^{\mu\nu} \cdot F_{\mu\nu} = - \frac {1}{4} {\tilde F}^{MN} \cdot {\tilde F}_{MN}, \label{l6}
\end{eqnarray}
where ${\tilde F}^{MN}$ is the super curvature defined on the $(3, 2)$-dimensional supermanifold and can be derived from 
the Maurer-Cartan equation: ${\tilde F}^{(2)} = \tilde d {\tilde A}^{(1)} + i g\;({\tilde A}^{(1)} \wedge {\tilde A}^{(1)}) \equiv  \frac {1}{2!} (dZ^M \wedge dZ^N){\tilde F}_{MN}$. 
The celebrated HC condition (\ref{l6}) leads to the following relationships amongst the basic, auxiliary and secondary fields
\begin{eqnarray}
&& R_\mu = D_\mu C, \qquad \bar R_\mu = D_\mu \bar C, \qquad B_1 = -  \frac{i}{2}\;g\;(C \times C), \qquad \bar B_2 = - \frac{i}{2}\; g\;(\bar C \times \bar C), \nonumber\\
&& S_\mu = D_\mu B \;+ \;i g \;(D_\mu C \times \bar C)\; \equiv\; - D_\mu \bar B - \;i g\;(D_\mu \bar C \times C), \nonumber\\
&& s = -g\;(\bar B \times C), \qquad \bar s = + g\;(B \times \bar C), \qquad B + \bar B = -i g\;(C \times \bar C),  \label{l7}
\end{eqnarray}
where we have made the choices $\bar B_1 = \bar B$ and $B_2 = B$ which are, finally, identified with the Nakanishi-Lautrup type auxiliary fields. Substituting these 
relationships in the superexpansion (\ref{l5}), we have following explicit expansions:
\begin{eqnarray}
\tilde B^{(h)}_{\mu} (x, \theta, \bar\theta) &=& A_\mu (x) + \theta\; D_\mu \bar C (x) 
+ \bar \theta\; D_\mu C (x) 
+  \theta \;\bar\theta \; [i D_\mu  B - g\;(D_\mu C \times \bar C)] (x) \nonumber\\
&\equiv& A_\mu (x) + \theta\; (s_{ab} A_\mu (x)) + \bar \theta\; (s_b A_\mu (x)) + \theta\; \bar\theta\;
(s_b s_{ab} \;A_\mu (x)), \nonumber\\
\tilde F^{(h)} (x, \theta, \bar\theta) &=& C (x) + \theta\; (i \bar B (x)) + \bar \theta\; \Bigl 
[\frac{g}{2}\; (C \times C) (x) \Bigr ]
+  \theta \bar\theta \; [-i g\;(\bar B \times C)(x)] \nonumber\\
&\equiv& C (x) + \theta\; (s_{ab} C (x)) + \bar \theta\; (s_b C (x)) + \theta\; \bar\theta\;
(s_b s_{ab} \;C (x)),\nonumber\\
\tilde {\bar F}^{(h)} (x, \theta, \bar\theta) &=& \bar C (x) +
\theta\; \Bigl [\frac{g}{2} \;(\bar C \times \bar C) (x) \Bigr ] + \bar \theta\; (i  B (x))
+ \theta \bar\theta \; [(+ ig\;(B \times \bar C) (x)] \nonumber\\
&\equiv& \bar C (x) + \theta\; (s_{ab} \bar C (x)) + \bar \theta\; (s_b \bar C (x)) + \theta\; \bar\theta\;  \label{l8}
(s_b s_{ab}\; \bar C (x)),
\end{eqnarray}
where $(h)$, as the superscript on the superfields, denotes the expansions of the superfields 
after the application of HC. Thus, we can read out the (anti-)BRST symmetry transformations ($s_{(a)b}$) corresponding to the gauge field $A_\mu$ and 
(anti-)ghost fields $(\bar C)C$ from the above expressions. The (anti-)BRST symmetry transformations corresponding to the auxiliary fields $B$ and $\bar B$
can be obtained from the requirement of nilpotency and absolute anticommutativity properties of (anti-)BRST symmetries. 

Furthermore, in order to derive the (anti-)BRST symmetry transformations for the vector field $\phi_\mu$ and the auxiliary field $\rho$, we have to go
beyond the HC. For this purpose, we take help of gauge invariant restrictions (GIR) constituted with the help of composite fields 
($F_{\mu\nu} \cdot \phi_\eta$) and ($F_{\mu\nu} \cdot \rho$) which remain invariant under gauge transformations (\ref{l2}). It is clear as below
\begin{eqnarray}
 \delta_1 (F_{\mu\nu}\cdot \phi_\eta) = 0, \qquad  \delta_1 (F_{\mu\nu} \cdot \rho) = 0. \label{l10}
\end{eqnarray}
These gauge invariant quantities are physical ones (in some sense), thus, they must remain unaffected by the presence of Grassmannian variables when 
former quantities are generalized onto the $(3, 2)$-dimensional supermanifold. Therefore, we have following GIR
\begin{eqnarray}
&&\tilde F_{\mu\nu}^{(h)} (x,\theta,\bar\theta) \cdot \tilde \phi_\eta (x,\theta,\bar\theta)
= F_{\mu\nu} (x) \cdot \phi_{\eta} (x) \nonumber\\
&&\tilde F_{\mu\nu}^{(h)} (x,\theta,\bar\theta) \cdot \tilde \rho (x,\theta,\bar\theta)
= F_{\mu\nu} (x) \cdot \rho (x).  \label{l11}
\end{eqnarray}
In the above,  $\tilde \phi_\mu (x,\theta,\bar\theta)$ and $\tilde \rho (x,\theta,\bar\theta)$ are superfields corresponding to the vector 
field $\phi_\mu (x)$ and $\rho (x)$, respectively, whereas $\tilde F^{(h)}_{\mu\nu} (x, \theta, \bar\theta)$ is super 2-form curvature tensor. 
Now, following the same procedure as outlined above, we find the (anti-)BRST symmetry transformations corresponding to vector field $\phi_\mu$ and 
auxiliary field $\rho$. In explicit form, these (anti-)BRST symmetry transformations are 
\begin{eqnarray}
&& s_{ab} A_\mu = D_\mu \bar C, \qquad s_{ab} \bar C = \frac {g}{2} \,(\bar C \times \bar C),  \qquad s_{ab} B = - g\,(B \times \bar C), \quad s_{ab} \bar B = 0,  \nonumber\\
&& s_{ab} \phi_\mu = - g\,(\phi_\mu \times \bar C), \qquad s_{ab} C = i \bar B, \qquad s_{ab} \rho = - g\,(\rho \times \bar C), \nonumber\\
&& s_b A_\mu = D_\mu C, \qquad s_b C =  \frac{g}{2}\; (C \times C), \qquad s_b \bar B = - g\,(\bar B \times C), \quad s_b B = 0,  \nonumber\\
&& s_b \phi_\mu = - g\,(\phi_\mu \times C), \qquad s_b \bar C = i B, \qquad s_b \rho = - g\, (\rho \times C). \label{l12}
\end{eqnarray} 
Furthermore, it can be checked that the above mentioned (anti-)BRST symmetry transformations are off-shell nilpotent 
(i.e. $s_{(a)b}^2 = 0$) and absolutely anticommunting (i.e. $s_b s_{ab} + s_{ab} s_b = 0$) in nature in their operator form.

\section{Curci-Ferrari Restriction} 
A close look at (\ref{l7}) reveals that the Curci-Ferrari restriction [$B + \bar B = -i g(C \times \bar C)$] is a natural outcome of superfield approach. Actually, 
this condition arises when we set $\tilde F_{\theta \bar\theta}$ component of supercurvature tensor to be zero. It connects the Nakanishi-Lautrup auxiliary fields 
$B$ and $\bar B$ with the (anti-)ghost fields $(\bar C)C$ of the theory. The CF restriction is a {\it hallmark} of any non-Abelian 1-form gauge theory \cite{cur} 
and plays a central role in providing the proof for absolute anticommutativity of (anti-)BRST symmetry transformations. It also plays an important role in obtaining 
a set of coupled Lagrangian densities which respect the above derived (anti-)BRST symmetry transformations (\ref{l12}). The details may be found in Ref. \cite {srm1,sgrk}.

\section{Conclusions} 
In this talk, we summarize our results on the $3D$ massive Jackiw-Pi model. We have derived (anti-)BRST symmetry transformations corresponding to the YM symmetries of 
JP model. One of the novel features of this investigation is the derivation of (anti-)BRST transformations for the auxiliary field $\rho$ from our superfield 
formalism which is {\it neither} generated by the (anti-)BRST charges {\it nor} obtained from the requirements of nilpotency and/or absolute anticommutativity of the 
(anti-)BRST symmetries for our 3D model. The Curci-Ferrari restriction, which plays a central role in providing the proof for absolute anticommutativity of (anti-)BRST 
symmetry transformations, is a natural outcome of this superfield approach. \\

\noindent
{\bf Acknowledgments:} Financial support from The Institute of Mathematical Sciences, Chennai, India is gratefully acknowledged. The author would also like to 
thankfully acknowledge his collaborators, R. P. Malik and R. Kumar, with whom the present work is completed.

\end{document}